\documentclass[twocolumn,epsfig,superscriptaddress,prl]{revtex4-2}

\usepackage{graphicx}
\usepackage{amssymb}
\usepackage{amsmath}
\usepackage{hyperref}
\usepackage{changes}
\usepackage{comment}

\newcommand{\tr}{\text{tr}}
\newcommand{\adj}{\text{adj}}

\begin{document}
\title{Stacking-Induced Symmetry-Protected Topological Phase Transitions}	

\author{Sang-Jun Choi}
\email{sang-jun.choi@physik.uni-wuerzburg.de}
\affiliation{Institute for Theoretical Physics and Astrophysics, University of W\"urzburg, D-97074 W\"urzburg, Germany}
\author{Bj\"orn Trauzettel}
\affiliation{Institute for Theoretical Physics and Astrophysics, University of W\"urzburg, D-97074 W\"urzburg, Germany}
\affiliation{W\"urzburg-Dresden Cluster of Excellence ct.qmat, Germany}
\date{\today}

\begin{abstract}
{
We study symmetry-protected topological (SPT) phase transitions induced by stacking two gapped one-dimensional subsystems in BDI symmetry class. The topological invariant of the entire system is a sum of three topological invariants: two from each subsystem and an emerging topological invariant from the stacking. We find that any symmetry-preserving stacking of topologically trivial subsystems can drive the entire system into a topologically nontrivial phase for a certain coupling strength. We explain this intriguing SPT phase transitions by conditions set by orbital degrees of freedom and time-reversal symmetry. To exemplify the SPT transition, we provide a concrete model which consists of an atomic chain and a spinful nanowire with spin-orbit interaction and $s$-wave superconducting order. The stacking-induced SPT transition drives this heterostructure into a zero-field topological superconducting phase.
}
\end{abstract}
\maketitle

The search for exotic topological excitations in symmetry-protected topological (SPT) phases has been extensively pursued in the last decades. The classic examples of SPT phases have become time-reversal symmetric topological insulators~\cite{Schnyder,Ludwig,Kitaev2009,Hasan,Ando}. Subsequently, SPT phases have been enriched by considerations of space group symmetries~\cite{Fu,Kruthoff}, a complete topological electronic band theory~\cite{Bradlyn}, and higher-order topological insulators~\cite{Schindler,Xue}. In addition to material realizations, quasicrystals~\cite{Oded,Fan}, synthetic dimensions~\cite{Yuan,Ozawa}, stacking layers of materials~\cite{Paolo,Budich,Baum}, and non-Hermitian systems~\cite{Bergholtz} have been studied as platforms hosting SPT phases.

Moreover, SPT phases are connected to unconventional superconductors. Kitaev has proposed in a seminal work a toy model stabilizing a one-dimensional topological superconductor (TSC)~\cite{Kitaev2001}, which is a superconducting SPT phase~\cite{Nayak}.
This model is a paradigmatic example, which has sparked a new research area regarding heterostructures hosting SPT phases~\cite{Alicea, Beenakker, Stanescu2011, Stanescu2013, Kane, Lutchyn, Oreg, Mourik, Nadj, Deng}. However, most proposals realizing SPT phases require a particular type of interaction.

While it is straightforward to analyze whether a SPT phase is allowed for a combination of symmetries ~\cite{Schnyder,Kitaev2009}, it is nontrivial to identify interactions driving systems into SPT phases. We address this point from a rather general perspective by stacking two SPT systems.  Interestingly, we find that any symmetry-preserving stacking of topologically trivial subsystems exhibits a nontrivial SPT phase. This SPT phase transition relies on orbital degrees of freedom (ODF) of the two subsystems and symmetries of the BDI class. Hence, ODF can be a crucial element of nontrivial SPT phases complementing the search for specific interactions.

In this work, we study SPT phase transitions induced by stacking two one-dimensional subsystems with arbitrary but symmetry-preserving coupling. Focusing on BDI symmetry class, we provide a general form of the Hamiltonian combining two subsystems with ODF $2N_u$ and $2N_d$, respectively. We derive the winding number of the entire system as a topological invariant. We show that the winding number includes a quantized winding number emerging from the stacking in addition to two individual winding numbers from each subsystem. Remarkably, any symmetry-preserving coupling drives two topological trivial subsystems into a nontrivial SPT phase at a sufficient coupling strength, provided that $(-1)^{\min\{N_u,N_d\}}=-1$. The stacking-induced SPT transition occurs as time-reversal symmetry obstructs the topologically trivial ground state manifold of two subsystems to deform continuously to another trivial manifold with increasing coupling strength for a given set of ODF.

We provide a concrete model which consists of an atomic chain and a spinful nanowire with spin-orbit interaction and $s$-wave superconducting order. The atomic chain is topological trivial and the same is true for the spinful nanowire without a magnetic field. We show that stacking those two systems induces a nontrivial SPT superconducting phase. This results in a zero-field TSC hosting Majorana bound states. 

We identify a physical mechanism responsible for the formation of a one-dimensional helical state. In the pioneering works, a magnetic field has been proposed to lift Kramer's degeneracy and mimic helicity~\cite{Oreg, Lutchyn, Mourik, Deng}. In our model, the symmetry-preserving stacking prevents the system from being symmetric with respect to the time-reversal operator $\hat{T}$ satisfying $\hat{T}^2=-\hat{\mathbf{1}}$. Hence, the stacking lifts Kramer's degeneracy and yields a helical state.

{\it General argument.}---We derive a general matrix representation $\mathcal{H}$ of the Hamiltonian $\hat{H}$ describing a stack of two one-dimensional BDI systems with Hamiltonians $\hat{H}_u$ and $\hat{H}_d$ and an arbitrary coupling term $\hat{Z}$. In momentum space, we analyze
\begin{equation}
\mathcal{H}(k) = \left(
\begin{array}{cc}
\mathcal{H}_u(k) & \mathcal{Z}^\dagger \\
\mathcal{Z} & \mathcal{H}_d(k)
\end{array}
\right).
\end{equation}
Since BDI symmetry class includes chiral symmetry, chiral operators $\mathcal{S}_\mu$ exist satisfying $\{\mathcal{H}_\mu(k), \mathcal{S}_\mu\}=0$ and $\mathcal{S}_\mu^2=\mathbf{1}$ ($\mu=u,d$). Note that the ODF of each subsystem are even integers $2N_u$ and $2N_d$ according to the eigenvalues $\pm1$ of chiral operators.
Using the eigenstates of the chiral operator for each block, we can transform the Hamiltonian into
\begin{equation}
\mathcal{H}(k) = \left(
\begin{array}{cc|cc}
\mathbf{0}_{N_u} & h_u^\dagger(k) & V^\dagger & Y^\dagger \\
h_u(k) & \mathbf{0}_{N_u} & X^\dagger & W^\dagger \\
\hline
V & X  & \mathbf{0}_{N_d} & h_d^\dagger(k) \\
Y & W & h_d(k) & \mathbf{0}_{N_d} \\
\end{array}
\right).
\end{equation}
The matrix elements of the coupling term $\mathcal{Z}$ are chosen as $V,W,X,Y$.

Since the entire system shall respect the chiral symmetry, we can reduce matrix elements further. The chiral symmetry operator $\mathcal{S}$ for the entire system is
\begin{equation}
\mathcal{S}=\left(
\begin{array}{cc}
\mathcal{S}_u & \mathbf{0} \\
\mathbf{0} & \mathcal{S}_d
\end{array}
\right).  \label{Eq:S}
\end{equation}
From $\{\mathcal{H}(k),\mathcal{S}\}=0$, we find the symmetry-preserving coupling should satisfy $X=Y=0$. By basis transformation, we can write the full Hamiltonian in the convenient form
\begin{equation}
\mathcal{H}(k) = \left(\begin{array}{cc}
\mathbf{0} & h^\dagger(k) \\
h(k) & \mathbf{0}
\end{array}
\right) ,\,\,
h(k) = \left(
\begin{array}{cc}
h_u(k) & W^\dagger \\
V & h_d(k)
\end{array}
\right). \label{Eq:H(k)}
\end{equation}
The matrix size of $h(k)$ is $N_u+N_d$. The BDI class includes time-reversal symmetry $\hat{T}$ satisfying $\hat{T}^2=\hat{\mathbf{1}}$ so that $H^*(x)=H(x)$~\cite{Haake}, which is equivalent to $\mathcal{H}^*(k)=\mathcal{H}(-k)$ in momentum space~\cite{Schnyder,Ludwig}. Consequently, $\mathcal{H}(k)$, $V$, and $W^\dagger$ are real matrices at the time-reversal momenta $k=0$ and $k=\pi$.

We calculate the winding number $\mathcal{W}$ of the entire system to study SPT phase transitions. Note that $h(k)$ can be continuously deformed except for singular points $z(k)\equiv\det[h(k)]=0$, at which the system is gapless. Hence, $z(k)$ can be viewed as a mapping of a closed loop $S^1$ to another closed loop $S^1$ in the complex plane, since $k\in S^1$ and $z(k)=z(k+2\pi)$. This defines the fundamental group of $\pi_1(S^1)\cong\mathbb{Z}$, and the number of times that $z(k)$ winds around $z(k)=0$ is given by
\begin{equation}
\mathcal{W} = \frac{1}{2\pi i}\int_0^{2\pi}dk \frac{\partial}{\partial k} \log\left\{\det\left[h(k)\right]\right\}.
\end{equation}
We find that $\mathcal{W} = \mathcal{W}_u + \mathcal{W}_d + \mathcal{W}_{ud}$, where $\mathcal{W}_\mu = (i/2\pi)\int_0^{2\pi}dk \partial_k \log\left\{ \det\left[h_\mu(k)\right] \right\}$ is the winding number of each subsystem $\mu=u,d$. An additional winding number $\mathcal{W}_{ud}$ emerges from stacking the two subsystems. It can be written as
\begin{equation}
\mathcal{W}_{ud} 
   = \frac{1}{2\pi i}\int_0^{2\pi}dk \frac{\partial \log\left\{\det\left[\mathbf{1}- h_d^{-1}(k)V h_u^{-1}(k) W^\dagger\right]\right\} }{\partial k}. \label{Eq:Wud}
\end{equation}
We emphasize that if we stack two topologically trivial subsystems, $\mathcal{W}_u=\mathcal{W}_d=0$, nontrivial SPT phase transitions can occur due to the emerging quantized winding number $\mathcal{W}_{ud}$. Surprisingly, the whole system of two trivial subsystems generically undergoes the SPT phase transitions for any symmetry-preserving stacking at sufficiently strong coupling, if $(-1)^{\min\{N_u,N_d\}}=-1$.

\begin{figure}[b]
\centering
\includegraphics[width=0.99\columnwidth]{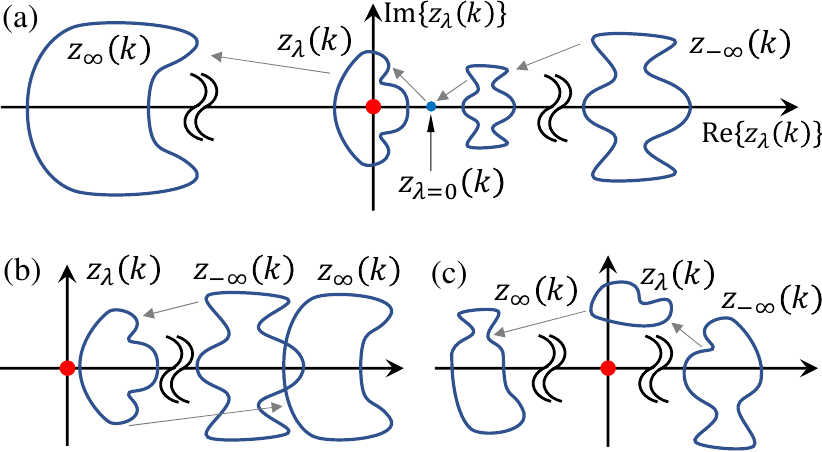}
\caption{
Symmetry-preserving stacking of two trivial subsystems induces a nontrivial SPT phase with nonzero winding number $\mathcal{W}_{ud}$ evaluated by the loop $z_\lambda(k)$ (blue). The loops deform by tuning the coupling strength $\lambda$. (a) Deformation of the loop in presence of time-reversal symmetry and $(-1)^{\min\{N_u,N_d\}}=-1$. (b) Same as (a) for $(-1)^{\min\{N_u,N_d\}}=+1$. (c) Same as (a) in absence of time-reversal symmetry.
} \label{Fig1}
\end{figure}

We first illustrate the SPT phase transition geometrically by analyzing the winding number $\mathcal{W}_{ud}$ as a function of coupling strength $\lambda$. We define a loop $z_\lambda(k)\equiv\det\left[\mathbf{1}- \lambda h_d^{-1}(k)V h_u^{-1}(k) W^\dagger\right]$ in the complex plane. The winding number $\mathcal{W}_{ud}$ counts the number of times that the loop $z_\lambda(k)$ winds around the origin with varying coupling strength $\lambda$~(see Fig.~\ref{Fig1}). The ODF of a subsystem let the loop deform from a loop in the far right (left) complex plane to a loop in the far left (right) complex plane, as sending $\lambda$ from $-\infty$ to $\infty$. In the limits $\lambda\rightarrow\pm\infty$, the loop becomes $z_\lambda(k) \propto (-\lambda)^{\min\{N_u,N_d\}}$~\cite{footnote2}. If $(-1)^{\min\{N_u,N_d\}}=-1$, the loop is found at $z_\lambda(k)\rightarrow \infty$ as $\lambda\rightarrow-\infty$ and at the other side $z_\lambda(k)\rightarrow -\infty$ as $\lambda\rightarrow\infty$, or vice versa. Time-reversal symmetry sets conditions on the deformation of the loops as we change $\lambda$ between the limits. It forces the loops to be symmetrical about the real axis, i.e., $z^*_\lambda(k) = z_\lambda(2\pi-k)$. Additionally, it causes the loops to intersect with the real axis at the time-reversal invariant momenta of $k=0$ and $k=\pi$. Hence, when $\min\{N_u,N_d\}$ is odd, time-reversal symmetry obstructs two loops $z_\infty(k)$ and $z_{-\infty}(k)$ to continuously deform into each another without passing the origin [see Fig.~\ref{Fig1}(a)]. Thus, any symmetry-preserving stacking drives the combined system into nontrivial SPT phases with nonzero $\mathcal{W}_{ud}$ for certain coupling strengths. However, if $\min\{N_u,N_d\}$ is even, the continuous deformation by $\lambda$ does not guarantee a nonzero $\mathcal{W}_{ud}$ [Fig.~\ref{Fig1}(b)]. Likewise, if time-reversal symmetry is broken, the loop is able to circumvent the origin [Fig.~\ref{Fig1}(c)].

\begin{figure}[t]
\centering
\includegraphics[width=0.99\columnwidth]{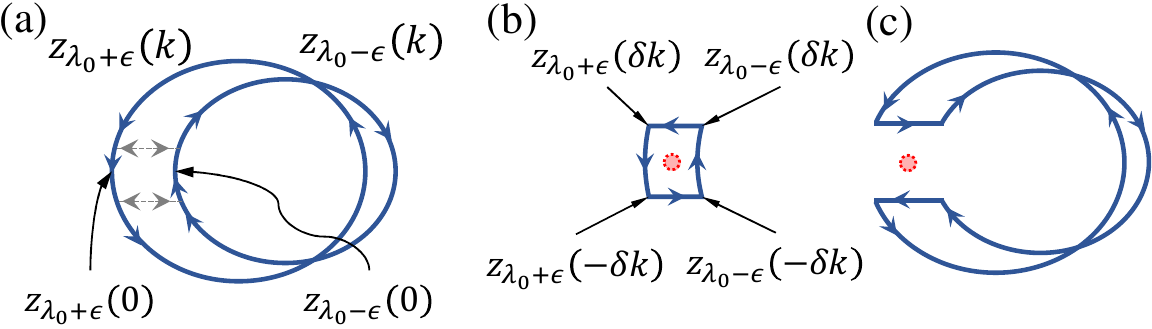}
\caption{(a) Two loops of $z_{\lambda_0+\epsilon}(k)$ and $z_{\lambda_0-\epsilon}(k)$ yield $\delta\mathcal{W}_{ud}$: The sum of contour integrals along the loops shown in (b) and (c) equals $\delta\mathcal{W}_{ud}$.
} \label{Fig2}
\end{figure}

We support the above geometric picture of the SPT phase transition by studying analytical properties of the winding number $\mathcal{W}_{ud}$. Note that the necessary condition for a SPT phase transition is a gap closing: $z_\lambda(k)=\det\left[\mathbf{1}- \lambda h_d^{-1}(k)V h_u^{-1}(k) W^\dagger\right]=0$ for a real-valued coupling strength $\lambda$. Due to time-reversal symmetry, $z_\lambda(k)=0$ becomes a polynomial equation of the variable $\lambda$ with real coefficients when $k=0$ and $k=\pi$. According to the fundamental theorem of algebra, there exists at least one real solution $\lambda_0$ if the degree of the polynomial equations is odd~\cite{footnote2}. This condition is satisfied for an odd integer $\min\{N_u,N_d\}$. The gap closing occurs for a certain symmetry-preserving coupling strength. 

Since the gap closing alone does not guarantee SPT phase transitions, we show that the winding number changes after the gap closing: $\delta\mathcal{W}_{ud}=\lim_{\epsilon\rightarrow0}[\mathcal{W}_{ud}(\lambda_0+\epsilon)-\mathcal{W}_{ud}(\lambda_0-\epsilon)]\ne0$ [see Fig.~\ref{Fig2}(a)]. We prove a nonzero $\delta\mathcal{W}_{ud}$ on the basis of the loops in Fig.~\ref{Fig2}(b) and (c). Due to time-reversal symmetry, we find $z_{\lambda_0+\epsilon}(\delta k)=-x\epsilon + iy\delta k$ with nonzero real numbers $x$ and $y$ close to a gap closing momentum $k=0$~\cite{SM}, where $\epsilon$ and $\delta k$ are infinitesimally small. We derive the positions of the other vertices of the loop in Fig.~\ref{Fig2}(b) in the complex plane as $z_{\lambda_0+\epsilon}(-\delta k) =   -x\epsilon - iy\delta k$, $z_{\lambda_0-\epsilon}(-\delta k) = x\epsilon - iy\delta k$, and $z_{\lambda_0-\epsilon}(\delta k) = x\epsilon + iy\delta k$. Since $x$ and $y$ are nonzero, the loop in Fig.~\ref{Fig2}(b) encloses the origin with a nonzero winding number. The winding number of the loop in Fig.~\ref{Fig2}(c) is zero instead.

\begin{figure}[b]
\centering
\includegraphics[width=0.99\columnwidth]{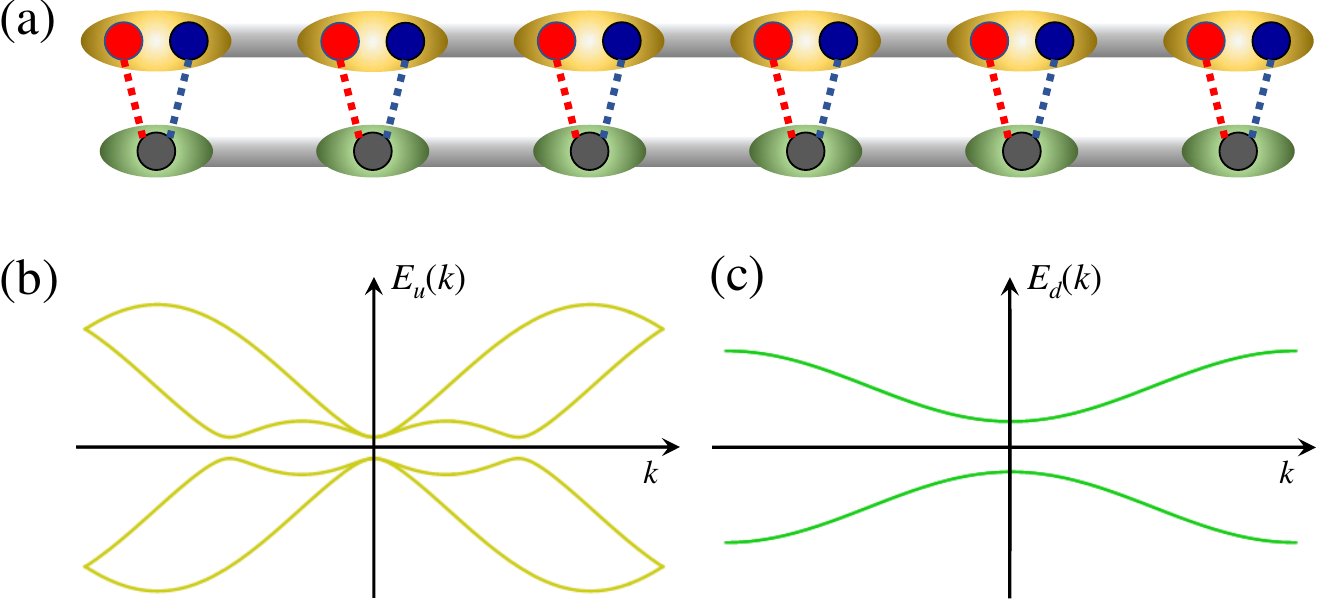}
\caption{Stacking of two topologically trivial subsystems yields a zero-field TSC phase. (a) Spinful (top) and spin-polarized (bottom) quantum wires are stacked with arbitrary coupling strength (dashed lines). (b) The spinful wire exhibits an $s$-wave superconducting gap along with spin-orbit coupling. (c) The chemical potential gaps out the spin-polarized wire. The lower branchs $E_{u,d}<0$ depict the hole band.
} \label{Fig3}
\end{figure}

{\it Zero-field TSC.}---We now present a specific example for the general result on the stacking-induced SPT phase transition. We consider a heterostructure formed by stacking two topologically trivial one-dimensional subsystems (Fig.~\ref{Fig3}). One of the subsystems resembles of a semiconductor nanowire with spin-orbit coupling and $s$-wave superconducting order in absence of an external magnetic field. We show that stacking another spin-polarized wire to the first one induces a nontrivial SPT phase, resulting in zero-field TSC without {\it external} magnetic fields.

The Hamiltonian of the heterostructure is $\hat{H}_\text{BdG} = \hat{H}_u + \hat{H}_d + \hat{Z} + \hat{Z}^\dagger$. $\hat{H}_u = \int dk [\hat{\psi}^\dagger(k) \mathcal{H}_u(k) \hat{\psi}(k)]$ is the standard Hamiltonian of semiconductors with the spin-orbit coupling and superconducting order, regarding real materials such as InAs and InSb~\cite{Stanescu2011,Stanescu2013,Lutchyn}, $\hat{H}_d = \int dk [\hat{\phi}^\dagger(k) \mathcal{H}_d(k) \hat{\phi}(k)]$ is the Hamiltonian describing spin-polarized wire, and $\hat{Z} = \int dk \hat{\psi}^\dagger(k) \mathcal{Z} \hat{\phi}(k)$ is the coupling Hamiltonian,
\begin{eqnarray}
\mathcal{H}_u &=& \tau_z\left[t_u\cos k - \mu_u  + l_\text{so} \sin k \,\sigma_y \right] + \Delta\tau_x, \nonumber\\
\mathcal{H}_d &=& \tau_z\left[t_d\cos k -\mu_d \right], \nonumber\\
\mathcal{Z} &=& \left(
\begin{array}{cccc}
\frac{-v_2-w_2}{2} & \frac{v_1+w_1}{2} & \frac{-v_1+w_1}{2} & \frac{-v_2+w_2}{2} \\
\frac{-v_2+w_2}{2} & \frac{v_1-w_1}{2} & \frac{-v_1-w_1}{2} & \frac{-v_2-w_2}{2}
\end{array}\right). \label{Eq:Hstack}
\end{eqnarray}
$\hat{\psi}^\dagger(k)=[\hat{c}^\dagger_\uparrow(k) ,\, \hat{c}^\dagger_\downarrow(k) ,\, \hat{c}_\downarrow(k) ,\, -\hat{c}_\uparrow(k)]$ and $\hat{\phi}^\dagger(k) = [\hat{d}^\dagger(k) ,\, \hat{d}(k)]$. Here, $\hat{c}_{\uparrow(\downarrow)}(k)$ annihilates spin-up (down) electrons with momentum $k$ in the spinful wire, while $\hat{d}(k)$ annihilates electrons with momentum $k$ in the spin-polarized wire. The Pauli matrices $\tau$ and $\sigma$ act on particle-hole and spin space, respectively. The spin-orbit coupling and superconducting order of the spinful wire are described with real parameters $l_\text{so}$ and $\Delta$, respectively. The hopping energy and chemical potential of each subsystem $\mu=u,d$ are given by $t_\mu$ and $\mu_\mu$. Finally, the arbitrary symmetry-preserving coupling $\mathcal{Z}$ of both subsystems has four independent real parameters $v_1, v_2, w_1, w_2$. The matrix elements in the second row  of $\mathcal{Z}$ follow from imposing particle-hole symmetry. We note that the two subsystems alone remain topologically trivial for any set of physical parameters. 

The heterostructure satisfies the right conditions for the stacking-induced SPT phase transition: ODFs of subsystems $N_u$ and $N_d$ yield $(-1)^{\min\{N_u,N_d\}}=-1$, and the stacked system possesses the symmetries of BDI class.
Evidently, the first condition is satisfied, as $N_u=2$ and $N_d=1$ [see Eq.~\eqref{Eq:Hstack}]. Moreover, the heterostructure falls into the BDI symmetry class. The chiral operators of each subsystem are $\mathcal{S}_u = \tau_y\sigma_y$ and $\mathcal{S}_d=\tau_x$. We find $\{\hat{H}_\text{BdG},\hat{S}\}=0$ using the chiral operator for the whole system as in Eq.~\eqref{Eq:S}~\cite{SM}. We use the complex conjugate $\mathcal{K}$ as the time-reversal operator $\hat{T}$ satisfying $\hat{T}\hat{H}_\text{BdG}(k)\hat{T}^{-1} = \hat{H}_\text{BdG}(-k)$ with $\hat{T}^2=\hat{\mathbf{1}}$. The particle-hole symmetry operator is given by $\Xi=\mathcal{S}\mathcal{T}$.

\begin{figure}[t]
\centering
\includegraphics[width=0.99\columnwidth]{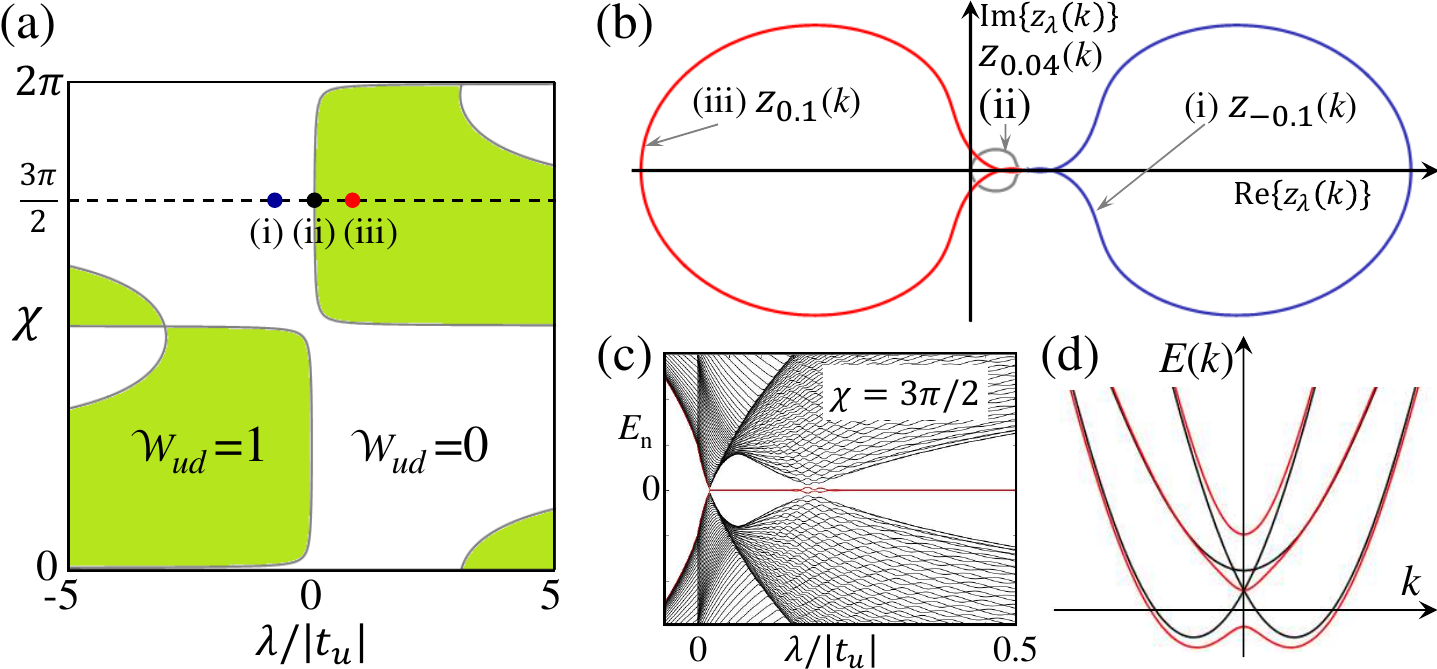}
\caption{(a) Phase diagram of the zero-field TSC (green) in terms of the coupling angle $\chi$ and strength $\lambda$. (b) $\mathcal{W}_{ud}=\int_0^{2\pi}dk\arg\{z_\lambda(k)\}/(2\pi)$ counts the number of times the loop $z_\lambda(k)$ winds around the origin. The loops depict the cases (i)-(iii) in the panel (a). (c) Energy spectra of the heterostructure with a finite size (200 unit cells). The Majorana bound states appear at zero energy (red). (d) Electron bands for $\lambda=0$ (black) and $\lambda\ne0$ (red). We obtain all figures with $t_u=-1, t_d=-0.7, \mu_u=-1.05, \mu_d=-0.8, l_\text{so}=0.5, \Delta=0.3$.
} \label{Fig4}
\end{figure}

We identify the stacking-induced SPT transition of the heterostructure with the winding number $\mathcal{W}_{ud}$ given in Eq.~\eqref{Eq:Wud}. The block off-diagonal representation of the Hamiltonian of the whole system as in Eq.~\eqref{Eq:H(k)} can be written as
\begin{equation}
h(k) = \left(
\begin{array}{ccc}
t_u\cos k -\mu_u & -\Delta - il_\text{so}\sin k & w_1 \\
\Delta + il_\text{so}\sin k & t_u\cos k -\mu_u & w_2 \\
v_1 & v_2 & t_d\cos k -\mu_d
\end{array}
\right).
\end{equation}
From this expression, the loop $z_{\lambda}(k)=\det[\mathbf{1}-\lambda h^{-1}_d(k)Vh^{-1}_u(k)W^\dagger]$, yielding $\mathcal{W}_{ud}$, can be calculated. Phase boundaries at which the SPT transitions occur are found using $z_{\lambda}(k=0)=0$ and $z_{\lambda}(k=\pi)=0$ as follows:
\begin{equation}
\lambda=\frac{(t_d\pm\mu_d)[\Delta^2+(t_u\pm\mu_u)^2]}{(t_u\pm\mu_u)\cos\chi \pm \Delta\sin\chi},
\end{equation}
where we parameterize $v_1=\cos(\chi+\delta)$, $v_2=\sin(\chi+\delta)$, $w_1=\cos\delta$, and $w_2=\sin\delta$. The phase diagram depends on two parameters: coupling angle $\chi$ and coupling strength $\lambda$ [Fig.~\ref{Fig4}(a)]. We obtain the phase diagram from $\mathcal{W}_{ud}$ for regions surrounded by the phase boundaries [Fig.~\ref{Fig4}(b)]. 

We find Majorana bound states as a result of the SPT phase transition. Zero energy bound states appear in the energy spectra of a finite size system [Fig.~\ref{Fig4}(c)]. The Majorana wavefunction, which is particle-hole symmetric $\Xi\zeta_\pm(x)=\zeta_\pm(x)$, can be written as 
\begin{eqnarray}
\zeta_+(x) &\propto& (1 ,\, 1 ,\, 1 ,\, -1 ,\, \sqrt{2} ,\, \sqrt{2})^T e^{\frac{\sqrt{2}\Delta+\lambda}{\sqrt{2}l_\text{so}}x}, \nonumber\\
\zeta_-(x) &\propto& (i ,\, i ,\, -i ,\, i ,\, i\sqrt{2} ,\, -i\sqrt{2})^T e^{-\frac{\sqrt{2}\Delta+\lambda}{\sqrt{2}l_\text{so}}x}, \label{Eq:MajoranaWavefunction}
\end{eqnarray}
where $\chi=3\pi/2$, $\mu_{u,d} = t_{u,d}+\lambda/\sqrt{2}$. In Eq.~\eqref{Eq:MajoranaWavefunction}, a continuum version of the Hamiltonian of Eq.~\eqref{Eq:Hstack} is employed to simplify the expression. Defining $\hat{\Gamma}_\pm = \int dx\hat{\Psi}^\dagger(x)\zeta_\pm(x)$, we explicitly show that Majorana bound states appear since $\hat{\Gamma}_\pm^\dagger = \hat{\Gamma}_\pm$, where $\hat{\Psi}^\dagger(x)=[\hat{c}^\dagger_\uparrow(x) ,\, \hat{c}^\dagger_\downarrow(x) ,\, \hat{c}_\downarrow(x) ,\, -\hat{c}_\uparrow(x),\,\hat{d}^\dagger(x) ,\, \hat{d}(x)]$.

We elaborate on the physical mechanism of the zero-field TSC as imitating the one-dimensional helical state using ODF instead of  magnetic fields. It is known that the helical state stabilizes a TSC phase with an $s$-wave superconducting order~\cite{Kane}. To mimic the helical state, a magnetic field has been added to spin-orbit nanowires to lift Kramer's degeneracy and open a gap at $k=0$~\cite{Oreg, Lutchyn, Mourik, Deng}. The Kramer's degeneracy is often eliminated by adding a time-reversal symmetry breaking interaction $\hat{V}'$ so that $[\hat{T},\hat{H}+\hat{V}'] \ne 0$. However, it can also be lifted by making the system obeying time-reversal symmetry $[\hat{T},\hat{H}] = 0$ but with $\hat{T}^2=\hat{1}$. The latter is the case in our stacked system. Accordingly, the interaction among bands in our model opens a gap, lifting Kramer's degeneracy without a magnetic field [Fig.~\ref{Fig4}(d)]. 

To illustrate that the heterostructure mimics helical states, we provide a low-energy effective Hamiltonian of the electron band for weak coupling strength,
\begin{equation}
\mathcal{H}_\text{eff} = t_u\cos k - \mu_u + l_\text{so}\sin k \sigma_y + \frac{v^2(\mathbf{1} + \sigma_x)}{4(\mu_d - t_d\cos k)}, \label{Eq:Heff}
\end{equation}
where $\chi=3\pi/2$, $\delta=0$. $H_\text{eff}$ describes the spinful wire, since most electrons fill the spinful part below the Fermi energy. The last term stems from the stacking of the spin-polarized nanowire and breaks the physical time-reversal symmetry of the spinful nanowire $\mathcal{T}_u=i\sigma_y\mathcal{K}$. This replaces the Zeeman interaction from an external magnetic field. Hence, the heterostructure mimics one-dimensional helical states at the Fermi energy and realizes zero-field TSC with $s$-wave superconducting order. We emphasize that effective time-reversal breaking from stacking is momentum-dependent. Especially, if $|\mu_d|\sim|t_d|$, effective time-reversal breaking is maximized at the time-reversal invariant momentum $k=0$ and rapidly decreases at finite momenta. Hence, almost perfect helical states can be achieved in combination with a large energy gap at $k=0$. This is an advantage with respect to the formation of quasi-helical states in spin-orbit nanowires employing Zeeman interaction.

{\it Conclusion.}---Nontrivial SPT phases can be induced by any symmetry-preserving stacking of two subsystems in the one-dimensional BDI symmetry class. We demonstrate that the only requirement for SPT phase transitions is set on ODF of each subsystem  without particular types of stacking or interactions. We illustrate the general result on the stacking-induced SPT phase transitions with a concrete example of zero-field TSC. It consists of an atomic wire coupled to a spin-orbit nanowire with $s$-wave superconducting order. Stacking two systems induces a SPT transition so that the stacked systems becomes a TSC in absence of a magnetic field. Since zero-field TSC does not require a large $g$-factor, it can stimulate experimental searches for implementing TSC in a different range of materials.

\begin{acknowledgments}
We thank Jan C. Budich for stimulating discussions. This work was supported by the W\"urzburg-Dresden Cluster of Excellence on Complexity and Topology in Quantum Matter (EXC2147, project-id 390858490) and by the DFG (SFB1170 ``ToCoTronics''). We thank the Bavarian Ministry of Economic Affairs, Regional Development and Energy for financial support within the High-Tech Agenda Project ``Bausteine f\"ur das Quanten Computing auf Basis topologischer Materialen''.
\end{acknowledgments}

\pagebreak
\widetext
\begin{center}
\textbf{\large Supplemental Material: Stacking-induced Symmetry-protected Topological Phase Transitions }
\end{center}

\section{Derivation for $z_{\lambda_0+\epsilon}(\delta k)  = -x\epsilon + iy\delta k$}
We provide the detail of the derivation of $z_{\lambda_0+\epsilon}(\delta k) = -x\epsilon + iy\delta k$ in this supplemental material. Applying the Jacobi's formula to $z_{\lambda}(k) = \det\left[\mathbf{1}- \lambda h_d^{-1}(k)V h_u^{-1}(k) W^\dagger\right]$ with the infinitesimally small $\epsilon$ and $\delta k$, we obtain
\begin{eqnarray}
z_{\lambda_0+\epsilon}(\delta k) & \simeq & -\tr\{\adj[\mathbf{1}-\lambda_0 h_d^{-1}(0)V h_u^{-1}(0)W^\dagger] \, h_d^{-1}(0)V h_u^{-1}(0)W^\dagger \}\epsilon \nonumber\\
&& -\tr\{\adj[\mathbf{1}-\lambda_0 h_d^{-1}(0)V h_u^{-1}(0)W^\dagger]\,\partial_k[h_d^{-1}(k)V h_u^{-1}(k)W^\dagger]|_{k=0} \}\delta k \label{Eq:z_uv1}\\
&=&  -x\epsilon + iy\delta k, \label{Eq:z_uv2}
\end{eqnarray}
where the zeroth order of $\epsilon$ and $\delta k$ vanishes, as $\det[\mathbf{1}-\lambda_0 h_d^{-1}(0)V h_u^{-1}(0)W^\dagger]=0$. Since $[z_{\lambda_0+\epsilon}(\delta k)]^* =  z_{\lambda_0+\epsilon}(2\pi-\delta k) =  z_{\lambda_0+\epsilon}(-\delta k)$ due to the time-reversal symmetry and $2\pi$-periodic $z_\lambda(k)$, the prefactor of $\delta k$ in Eq.~\eqref{Eq:z_uv1} is purely imaginary, while that of $\epsilon$ is real. Using Eq.~\eqref{Eq:z_uv2} with some real numbers of $x$ and $y$, we get
\begin{eqnarray}
z_{\lambda_0+\epsilon}(\delta k)  & \simeq &  -x\epsilon + iy\delta k, \\
z_{\lambda_0+\epsilon}(2\pi-\delta k) & \simeq &   -x\epsilon - iy\delta k, \\
z_{\lambda_0-\epsilon}(2\pi-\delta k) & \simeq & x\epsilon - iy\delta k, \\
z_{\lambda_0-\epsilon}(\delta k) & \simeq & x\epsilon + iy\delta k,
\end{eqnarray}
We now show that $x$ and $y$ are nonzero. We introduce a shorthand notation $h_0\equiv\lambda_0 h_d^{-1}(0)V h_u^{-1}(0)W^\dagger$ for convenience. Since $[\adj(\mathbf{1}-h_0)](\mathbf{1}-h_0)=\det(\mathbf{1}-h_0)\mathbf{1}=0$, we simplify $x = \tr[\adj(\mathbf{1}-h_0)]$. Using Jacobi's formula, we write
\begin{eqnarray}
x = \tr\left[ \adj(\rho\mathbf{1}-h_0)\frac{\partial (\rho\mathbf{1}-h_0)}{\partial\rho} \right]_{\rho=1}
  = \left[\frac{\partial\det(\rho\mathbf{1}-h_0)}{\partial\rho} \right]_{\rho=1}.
\end{eqnarray}
Notice that $\det(\rho\mathbf{1}-h_0)$ is the characteristic polynomial $C(\rho)$ of $h_0$. Since $\det(\mathbf{1}-h_0)=0$ and $h_0$ is a real $m\times m$-matrix, $C(\rho)$ is factorized into $C(\rho)=(\rho-1)(\rho-w_1)\cdots(\rho-w_{m-1})$.
\begin{eqnarray}
x &=& \left[\frac{\partial\det(\rho\mathbf{1}-h_0)}{\partial\rho} \right]_{\rho=1} \nonumber\\
 &=& \left[\frac{\partial}{\partial\rho} (\rho-1)(\rho-w_1)\cdots(\rho-w_{m-1}) \right]_{\rho=1} \nonumber\\
 &=& \left\{\{ (\rho-w_1)\cdots(\rho-w_{m-1})   + (\rho-1) \sum_{j=1}^{m-1}(\rho-w_1)\cdots(\rho-w_{j-1})(\rho-w_{j+1})\cdots(\rho-w_{m-1}) \right\}_{\rho=1} \nonumber\\
 &=& (1-w_1)\cdots(1-w_{m-1})\ne 0.
\end{eqnarray}
We exclude the particular case where $h_0$ are fine-tuned to have degenerated eigenvalues. As time-reversal symmetry enforces $[z_{\lambda_0+\epsilon}(\delta k)]^* = z_{\lambda_0+\epsilon}(-\delta k)$, the leading order of $\delta k$ becomes odd. Hence, $y$ is nonzero.

\section{Matrix representations $\mathcal{H}_\text{BdG}$ and $\mathcal{S}$ of the Hamiltonian $\hat{H}_\text{BdG}$ and chiral operator $\hat{S}$ of zero-field TSC}
We provide matrix representations $\mathcal{H}_\text{BdG}$ and $\mathcal{S}$ of the Hamiltonian $\hat{H}_\text{BdG}$ and the chiral operator $\hat{S}$ of zero-field TSC, respectively.
\begin{eqnarray}
\mathcal{H}_\text{BdG} &=& \left(
\begin{array}{cccccc}
t_u\cos k - \mu_u & -il_\text{so}\sin k & \Delta & 0 & \frac{-v_2-w_2}{2} & \frac{-v_2+w_2}{2} \\
il_\text{so}\sin k & t_u\cos k - \mu_u  & 0 & \Delta & \frac{v_1+w_1}{2} & \frac{v_1-w_1}{2} \\
\Delta & 0 & -t_u\cos k + \mu_u & il_\text{so}\sin k & \frac{-v_1+w_1}{2} & \frac{-v_1-w_1}{2} \\
0 & \Delta & -il_\text{so}\sin k & -t_u\cos k + \mu_u & \frac{-v_2+w_2}{2} & \frac{-v_2-w_2}{2} \\
\frac{-v_2-w_2}{2} & \frac{v_1+w_1}{2} & \frac{-v_1+w_1}{2} & \frac{-v_2+w_2}{2} & t_d\cos k-\mu_d & 0 \\
\frac{-v_2+w_2}{2} & \frac{v_1-w_1}{2} & \frac{-v_1-w_1}{2} & \frac{-v_2-w_2}{2} & 0 & -t_d\cos k+\mu_d
\end{array}\right),\\
\mathcal{S} &=& \left(
\begin{array}{cccccc}
0 & 0 & 0 & -1 & 0 & 0 \\
0 & 0 & 1 & 0 & 0 & 0 \\
0 & 1 & 0 & 0 & 0 & 0 \\
-1 & 0 & 0 & 0 & 0 & 0 \\
0 & 0 & 0 & 0 & 0 & 1 \\
0 & 0 & 0 & 0 & 1 & 0
\end{array}\right).
\end{eqnarray}


\begin{thebibliography}{70}
\bibitem{Schnyder} A. P. Schnyder, S. Ryu, A. Furusaki, and Andreas W. W. Ludwig, Classification of topological insulators and superconductors in three spatial dimensions, Phys. Rev. B {\bf78}, 195125 (2008).

\bibitem{Ludwig} Andreas W. W. Ludwig, Topological phases: classification of topological insulators and superconductors of non-interacting fermions, and beyond, Phys. Scr.{\bf2016}, 014001 (2016).

\bibitem{Kitaev2009} A. Kitaev, Periodic table for topological insulators and superconductors, AIP Conference Proceedings {\bf1134}, 22 (2009).


\bibitem{Hasan} M. Z. Hasan and C. L. Kane, Colloquium: Topological insulators, Rev. Mod. Phys. {\bf82}, 3045 (2010).


\bibitem{Ando} Y. Ando, Topological Insulator Materials, J. Phys. Soc. Jpn. {\bf82}, 102001 (2013).

\bibitem{Fu} Y. Ando and L. Fu, Topological crystalline insulators and topological superconductors, Annu. Rev. Condens. Matter Phys. {\bf6}, 361 (2015).

\bibitem{Kruthoff} Jorrit Kruthoff, Jan de Boer, Jasper van Wezel, Charles L. Kane, and Robert-Jan Slager, Topological Classification of Crystalline Insulators through Band Structure Combinatorics, Phys. Rev. X {\bf7}, 041069 (2017).

\bibitem{Bradlyn} Barry Bradlyn, L. Elcoro, Jennifer Cano, M. G. Vergniory, Zhijun Wang, C. Felser, M. I. Aroyo, and B. Andrei Bernevig, Topological quantum chemistry, Nature {\bf547}, 298 (2017).

\bibitem{Schindler} Frank Schindler, Ashley M. Cook, Maia G. Vergniory, Zhijun Wang, Stuart S. P. Parkin, B. Andrei Bernevig, Titus Neupert, Higher-order topological insulators, Sci. Adv. {\bf4}, eaat0346 (2018).

\bibitem{Xue} H. Xue, Y. Yang, F. Gao, Y. Chong, and B. Zhang, Acoustic higher-order topological insulator on a kagome lattice, Nat. Mater. {\bf18}, 108 (2019).

\bibitem{Oded} O. Zilberberg, Topology in quasicrystals, Optical Materials Express, {\bf11}, 1143 (2021).

\bibitem{Fan} J. Fan and H. Huang, topological states in quasicrystals, Front. Phys. {\bf17}, 13203 (2022).

\bibitem{Yuan} L. Yuan, Q. Lin, M. Xiao, and S. Fan, Synthetic dimension in photonics, Optica, {\bf5}, 1396  (2018).

\bibitem{Ozawa} T. Ozawa and Hannah M. Price, Topologcial quantum matter in synthetic dimensions, Nat. Rev. Phys. {\bf1}, 349 (2019).

\bibitem{Paolo} Paolo Michetti, Jan C. Budich, Elena G. Novik, and Patrik Recher, Tunable quantum spin Hall effect in double quantum wells, Phys. Rev. B {\bf85}, 125309 (2012).

\bibitem{Budich} Jan C. Budich, Bj\"orn Trauzettel, and Paolo Michetti, Time Reversal Symmetric Topological Exciton Condensate in Bilayer HgTe Quantum Wells, Phys. Rev. Lett. {\bf112}, 146405 (2014).

\bibitem{Baum} Yuval Baum, Thore Posske, Ion Cosma Fulga, Bj\"orn Trauzettel, and Ady Stern, Coexisting Edge States and Gapless Bulk in Topological States of Matter, Phys. Rev. Lett. {\bf114}, 136801 (2015).

\bibitem{Bergholtz} Emil J. Bergholtz, Jan Carl Budich, and Flore K. Kunst, Exceptional topology of non-Htermitian systems, Rev. Mod. Phys. {\bf93}, 015005 (2021).


\bibitem{Kitaev2001} A. Kitaev, Unpaired Majorana Fermions in Quantum Wires, Phys. Usp. {\bf44}, 131 (2001).

\bibitem{Nayak} C. Nayak, S. H. Simon, A. Stern, M. Freedman, and S. Das Sarma, Non-Abelian anyons and topological quantum computation, Rev. Mod. Phys. {\bf80}, 1083 (2008).

\bibitem{Alicea} J Alicea, New directions in the pursuit of Majorana fermions in solid state systems, Rep. Prog. Phys. {\bf75}, 076501 (2012).

\bibitem{Beenakker} C. W. J. Beenakker, Search for Majorana Fermions in Superconductors, Annu. Rev. Condens. Matter Phys. {\bf4}, 113 (2013).

\bibitem{Stanescu2011} T. D. Stanescu, R. M. Lutchyn, and S. D. Sarma, Majorana fermions in semiconductor nanowires, Phys. Rev. B {\bf84}, 144522 (2011).

\bibitem{Stanescu2013} T. D. Stanescu and S. Tewari, Majorana fermions in semiconductor nanowires: fundamentals, modeling, and experiment,  J. Phys.: Condens. Matter {\bf25}, 233201 (2013).

\bibitem{Kane} Liang Fu and C. L. Kane, Superconducting Proximity Effect and Majorana Fermions at the Surface of a Topological Insulator, Phys. Rev. Lett. {\bf100}, 096407 (2008).

\bibitem{Lutchyn} R. M. Lutchyn, J. D. Sau, and S. Das Sarma, Majorana Fermions and a Topological Phase Transition in Semiconductor-Superconductor Heterostructures, Phys. Rev. Lett. {\bf105}, 077001 (2010).

\bibitem{Oreg} Y. Oreg, G. Refael, and F. von Oppen, Helical Liquids and Majorana Bound States in Quantum Wires, Phys. Rev. Lett. {\bf105}, 177002 (2010).

\bibitem{Mourik} V. Mourik, K. Zuo, S. M. Frolov, S. R. Plissard, E. P. A. M. Bakkers, and L. P. Kouwenhoven, Signatures of Majorana fermions in hybrid superconductor-semiconductor nanowire devices, Science {\bf336}, 1003 (2012).

\bibitem{Deng} M. T. Deng, S. Vaitiekenas, E. B. Hansen, J. Danon, M. Leijnse, K. Flensberg, J. Nygård, P. Krogstrup, and C. M. Marcus, Majorana bound state in a coupled quantum-dot hybrid-nanowire system, Science {\bf354}, 1557 (2016).

\bibitem{Nadj} S. Nadj-Perge, I. K. Drozdov, J. Li, H. Chen, S. Jeon, J. Seo, A. H. MacDonald, B. A. Bernevig, and A. Yazdani, Observation of Majorana fermions in ferromagnetic atomic chains on a superconductor, Science {\bf346}, 602 (2014).



\bibitem{Haake} F. Haake, {\it Quantum Signatures of Chaos} (Springer, Berlin, 2010), Chap. 2, p. 20.

\bibitem{footnote2} The Sylvester's determinant theorem dictates that $\det(\mathbf{1}_{N_d} - \lambda h_d^{-1}V h_u^{-1} W^\dagger)=\det(\mathbf{1}_{N_u} - \lambda  h_u^{-1} W^\dagger h_d^{-1}V)$. Hence, the largest power of $\lambda$ in both polynomials is $\min\{N_u,N_d\}$. 


\bibitem{SM} See the Supplemental Material at (URL) for the derivation for $z_{\lambda_0+\epsilon}(\delta k)=-x\epsilon+iy\delta k$ and matrix representations $\mathcal{H}_\text{BdG}$ and $\mathcal{S}$ of the Hamiltonian $\hat{H}_\text{BdG}$ and chiral operator $\hat{S}$ of zero-field TSC.



\end{thebibliography}
\end{document}